\documentclass[12pt]{article}
\usepackage[russian,english]{babel}
\usepackage{amsfonts,longtable}
\usepackage{amsmath}
\usepackage{a4wide}
\usepackage{feynmf}
\begin{document}

\title{One-loop counterterms in the Yang-Mills theory with gauge invariant ghost field Lagrangian.}
\author{R. N. Baranov}
\date{}
\maketitle

\begin{abstract}
One-loop calculations of renormalization constants~$Z_1$, $Z_2$ in
the model, proposed in the paper ~\cite{Slavnov} with gauge invariant ghost field
Lagrangian are performed. It is shown that the model is asymptotically free and the renormalization constants satisfy the same equation
as in the ordinary Yang-Mills theory.
\end{abstract}

\section{Introduction.}
In the Yang-Mills theory ghost field Lagrangian doesn't possess
gauge symmetry and together with a gauge fixing term breaks gauge
invariance of the effective action. It makes problems for
regularization and renormalization of the theory. The
Slavnov-Taylor identities are more complicated then the Ward
identities in electrodynamics. Recently a new formulation of the
Yang-Mills theory was proposed ~\cite{Slavnov} in which  the gauge
invariance is broken only by the gauge fixing term. In the
paper~\cite{Slavnov} it was shown that in the framework of
perturbation theory this model is equivalent to the usual
Yang-Mills theory. Calculating the observables in this model one
can pass to the Lorentz-type gauge in which the renormalizability
is evident. Nevertheless a proof of renormalizability directly in
the gauge  proposed in the paper ~\cite{Slavnov} is absent.
\par In the present paper we calculate at one-loop the gauge field and
$A^3$-vertex renormalization constants. It is shown
that a relationship between this constants is the same as in the
ordinary Yang-Mills theory.
\par Let introduce notations. The Lagrangian proposed
in~\cite{Slavnov} looks as follows
\begin{equation}\label{L_Origin}
    \mathcal{L} = \mathcal{L}_{YM} + \left( D_\mu \varphi \right)^* \left( D_\mu \varphi
    \right) - \left( D_\mu X \right)^* \left( D_\mu X \right)
    + i \left( \left( D_\mu e \right)^* \left( D_\mu b \right) - \left( D_\mu
    b \right)^* \left( D_\mu e \right) \right)
\end{equation}
The fields~$\varphi$, $X$ are commuting complex $SU(2)$-doublets
and $e$, $b$ are anticommuting ones.
\begin{equation*}
    \begin{split}
    &\varphi = \frac{1}{\sqrt{2}} \begin{pmatrix} i\varphi^1 + \varphi^2 \\ \sqrt{2}g^{-1}a + \varphi^0 - i\varphi^3
    \end{pmatrix}; \qquad
    X = \frac{1}{\sqrt{2}} \begin{pmatrix} iX^1 + X^2 \\ -\sqrt{2}g^{-1}a + X^0 - iX^3
    \end{pmatrix};\\
    &e = \frac{1}{\sqrt{2}} \begin{pmatrix} ie^1 + e^2 \\ e^0 - ie^3
    \end{pmatrix}; \qquad
    b = \frac{1}{\sqrt{2}} \begin{pmatrix} ib^1 + b^2 \\ b^0 - ib^3 \end{pmatrix};
    \end{split}
\end{equation*}
 $D_\mu\varphi = \partial_\mu \varphi +
ig\frac{\tau^a}{2}A_\mu^a\,\varphi$ is the covariant derivative;  $\tau^a$ are Pauli matrices.
The Yang-Mills Lagrangian is $\mathcal{L}_{YM} =
-\frac{1}{4}F_{\mu\nu}^a F_{\mu\nu}^a$, $F_{\mu\nu}^a =
\partial_\nu A_\mu^a - \partial_\mu A_\nu^a + g\varepsilon^{abc} A_\mu^b
A_\nu^c$. Let introduce the fields $B^{\pm,a}=\varphi^a \pm X^a$,
$\varphi^\pm = \varphi^0 \pm X^0$. In terms of these fields the gauge transformation leaving  the Lagrangian~(\ref{L_Origin}) invariant is
\begin{equation}\label{gauge_transform}
    \left\{\begin{aligned}
    &\delta A_\mu^a(x) = -\partial_\mu \alpha^a(x) +
    g\varepsilon^{abc} A_\mu^b(x) \alpha^c(x)
    \\
    &\delta B^{+,a}(x) = \frac{g}{2} \varphi^+(x) \alpha^a(x) +
    \frac{g}{2}\varepsilon^{abc} B^{+,b}(x) \alpha^c(x)
    \\
    &\delta \varphi^+ = -\frac{g}{2} B^{+,a}(x) \alpha^a(x)
    \\
    &\delta B^{-,a}(x) = \sqrt{2}a \alpha^a(x) + \frac{g}{2} \varphi^-(x) \alpha^a(x) +
    \frac{g}{2}\varepsilon^{abc} B^{-,b}(x) \alpha^c(x)
    \\
    &\delta \varphi^- = -\frac{g}{2} B^{-,a}(x) \alpha^a(x)
    \\
    &\delta e^a(x) = \frac{g}{2} e^0(x) \alpha^a(x) +
    \frac{g}{2} \varepsilon^{abc} e^b(x) \alpha^c(x)
    \\
    &\delta e^0 = -\frac{g}{2} e^a(x) \alpha^a(x)
    \\
    &\delta b^a(x) = \frac{g}{2} b^0(x) \alpha^a(x) +
    \frac{g}{2} \varepsilon^{abc} b^b(x) \alpha^c(x)
    \\
    &\delta b^0 = -\frac{g}{2} b^a(x) \alpha^a(x)
    \end{aligned}\right.
\end{equation}
Being quantized in the Lorentz gauge $\partial_\mu A_\mu = 0$ the
model is described by the effective action which contains the
gauge fixing term and non gauge invariant Faddeev-Popov ghost
Lagrangian. However in this model the Lorentz invariant condition
$B^{-,a}=0$ is also an admissible gauge. In this case the gauge
invariance is broken by only the gauge fixing term:
$\mathcal{L}_{ef}=\mathcal{L}+\lambda^a B^{-,a}$, here
$\mathcal{L}$ is the Lagrangian~(\ref{L_Origin}). The Lagrangian
~(\ref{L_Origin}) at the surface $B^{-,a}=0$ looks as follows
\begin{multline}\label{L_B-is-zero}
    \mathcal{L} = \mathcal{L}_{YM} + \frac{1}{2} \partial_\mu
    \varphi^+ \partial_\nu \varphi^- + \frac{a}{\sqrt{2}} A_\mu^a
    \partial_\mu B^{+,a} + \frac{g}{4} A_\mu^a \left( \varphi^-
    \partial_\mu B^{+,a} - B^{+,a} \partial_\mu \varphi^- \right) +\\+
    \frac{ag\sqrt{2}}{4} A_\mu^2 \varphi^+ + \frac{g^2}{8} A_\mu^2
    \varphi^+ \varphi^- + i\left( \partial_\mu e^0 \partial_\mu
    b^0 + \partial_\mu e^a \partial_\mu b^a \right) +\\+ \frac{ig}{2}
    A_\mu^a \left\{ \varepsilon^{abc} (b^b \cdot \partial_\mu e^c -
    \partial_\mu b^b \cdot e^c) + (b^a \cdot \partial_\mu e^0 -
    \partial_\mu b^a \cdot e^0) - (b^0 \cdot \partial_\mu e^a -
    \partial_\mu b^0 \cdot e^a) \right\} +\\+ \frac{ig^2}{4} A_\mu^2
    (e^0 e^0 + e^a b^a)
\end{multline}
The propagators and vertices are given in the appendix.

\section{Renormalization constants.}
We start with considering the gauge field renormalization constant~$Z_2$ and
$A^3$-vertex renormalization constant $Z_1$ and show that in
the one-loop approximation these constants coincide with the
corresponding ones in the ordinary Yang-Mills theory. We also
consider anticommuting fields renormalization constant~$\bar{Z}_2$
and $A_\mu eb$-vertex renormalization constant~$\bar{Z}_1$ and show
 that the relationship $Z_1 Z_2^{-1} = \bar{Z}_1
\bar{Z}_2^{-1}$ holds in one-loop approximation. The
calculations can be found in the appendix. The dimensional
regularization is used, hence all tadpole diagrams are equal to zero.
\par The constant~$Z_2$ renormalizes the transversal part of one-particle irreducible diagrams with two
external gauge field lines~$A_\mu$. There are three types of the contributions
to the divergent part of the corresponding diagrams. The first one is a
contribution of gauge field~$A_\mu$ loops. This contribution
is the same as in the ordinary Yang-Mills
theory, because the vertices $A^3$, $A^4$ and the propagators coincide with the usual ones.
The second one is the contribution of loops with commuting scalar
fields~$B^{+,a}$, $\varphi^\pm$. And the third one is the contribution
of loops with anticommuting fields~$e$, $b$. The sum of the second
and the third contributions equals to
\begin{equation}\label{Z2_sumCommAnticomm}
    \frac{ig^2}{48\pi^2\varepsilon} \delta^{ab} (g_{\mu\nu}p^2 - p_\mu p_\nu)
\end{equation}
In the ordinary Yang-Mills theory divergent part of Faddeev-Popov
ghosts loop is the following~\cite{SlavnovFaddeev}, \cite{Peskin}.
\begin{equation}\label{Z2_FPGhosts}
    \frac{ig^2}{48\pi^2\varepsilon} \delta^{ab} (g_{\mu\nu}p^2 + 2p_\mu p_\nu)
\end{equation}
One can see that transversal parts of~(\ref{Z2_sumCommAnticomm})
and~(\ref{Z2_FPGhosts}) coincide. As it was mentioned above
the contributions of gauge field~$A_\mu$ loops in the present
theory and in the Yang-Mills theory coincide and therefore
constants~$Z_2$ coincide too.
\par Note that contrary to the standard formulation of the Yang-Mills theory the radiative corrections generate also renormalization of the longitudinal part of the polarization operator, that is the counterterm proportional to $(\partial_{\mu}A_{\mu})^2$. Such a counterterm does not break the gauge invariance, because it originates from the admissible gauge invariant counterterm $(D^2 \phi)^* ( D^2 \phi)$ after the shift $\phi^0 \rightarrow \phi^0+ \sqrt{2}g^{-1}a$.
\par The constant~$Z_1$ performs renormalization of
$A^3$-vertex. Again there are
three types of contributions. Contribution of the gauge field~$A_\mu$ loops
coincide with the corresponding one in the ordinary Yang-Mills
theory. In one-loop approximation the total contribution of
commuting scalar fields $B^{+,a}$, $\varphi^\pm$ and anticommuting
fields $e$, $b$ is equal to
\begin{equation}\label{Z1_sumCommAnticomm}
    \left.\left. \frac{g^3}{96\pi^2\varepsilon} \varepsilon^{abc}
    \right( g_{\mu\nu} (p-q)_\lambda - g_{\nu\lambda} (2p+q)_\mu +
    g_{\mu\lambda} (p+2q)_\nu \right),
\end{equation}
It coincides with the contribution of Faddeev-Popov ghosts in the
ordinary Yang-Mills theory and therefore the constant~$Z_1$ equals
Yang-Mills constant $Z_1$.
\par Thus in one-loop approximation the constants~$Z_2$, $Z_1$ are
the same as in the Yang-Mills theory and therefore the
expression~$Z_1 Z_2^{-1}$ is the same as in the Yang-Mills theory.
\begin{equation}
    Z_1 Z_2^{-1} = 1 -\frac{3g^2}{16\pi^2\varepsilon}
\end{equation}
which means that the model is asymptotically free.
\par To find the constants~$\bar{Z}_2$, $\bar{Z}_1$, it is necessary
to calculate the divergent part of the diagrams represented on
fig.~\ref{Fig_Z2Bar},~\ref{Fig_Z1Bar}. Performing the
calculations we obtain
\begin{gather}
    \bar{Z}_2 = 1 + \frac{9g^2}{32\pi^2\varepsilon}\nonumber\\
    \bar{Z}_1 = 1 + \frac{3g^2}{32\pi^2\varepsilon}\\
    \bar{Z_1} \bar{Z_2}^{-1} = \left( 1 +
    \frac{3g^2}{32\pi^2\varepsilon} \right) \left( 1 -
    \frac{9g^2}{32\pi^2\varepsilon} \right) = 1 + \left( \frac{3}{32}
    -\frac{9}{32} \right) \frac{g^2}{\pi^2\varepsilon} = 1
    -\frac{3g^2}{16\pi^2\varepsilon}\nonumber
\end{gather}
Thus $Z_1 Z_2^{-1} = \bar{Z_1} \bar{Z_2}^{-1}$.
\begin{figure}[h]
\center{
\begin{fmffile}{graphs5}
    \begin{fmfgraph*}(120,100)
    \fmfleft{x} \fmfright{y}
    \fmf{plain,tension=3,label.side=left,label=$e^0$}{x,x1} \fmf{phantom_arrow,tension=3,label.side=right,label=$p$}{x,x1}
    \fmf{plain,tension=3,label.side=left,label=$b^0$}{y1,y} \fmf{phantom_arrow,tension=3,label.side=right,label=$p$}{y1,y}
    \fmf{plain,label.side=right,label=$b^m$}{x1,xy} \fmf{plain,label.side=right,label=$e^n$}{xy,y1} \fmf{phantom_arrow,label.side=left,label=$p+k$}{x1,y1}
    \fmf{wiggly,right,label.side=right,label=$k$}{y1,x1}
    \fmfdot{x1,y1}
    \end{fmfgraph*}
\end{fmffile}

\begin{fmffile}{graphs6}
    \begin{fmfgraph*}(140,100)
    \fmfleft{xp,x,y} \fmfright{pp,p,z}
    \fmf{wiggly,tension=1.5,label.side=left,label=$p$}{y,x1} \fmf{phantom_arrow,tension=1.5}{y,x1}
    \fmf{plain,tension=3,label.side=left,label=$q$}{x,x1} \fmf{phantom_arrow,tension=3,label.side=right,label=$e^0$}{x,x1}
    \fmf{plain,tension=3,label.side=left,label=$q+p$}{z1,p} \fmf{phantom_arrow,tension=3,label.side=right,label=$b^b$}{z1,p}
    \fmf{phantom,tension=3}{z,z1}
    \fmf{plain,label.side=right,label=$b^0$}{x1,xz} \fmf{plain,label.side=right,label=$e^0$}{xz,z1} \fmf{phantom_arrow,label.side=left,label=$k+p+q$}{x1,z1}
    \fmf{wiggly,right,label.side=right,label=$k$}{z1,x1} \fmf{phantom_arrow,right}{z1,x1}
    \fmfdot{x1,z1}
    \fmflabel{$\mu\, a$}{y}
    \end{fmfgraph*}
    \begin{fmfgraph}(40,120)
    \end{fmfgraph}
    \begin{fmfgraph*}(140,100)
    \fmfleft{xp,x,y} \fmfright{pp,p,z}
    \fmf{wiggly,tension=1.5,label.side=right,label=$p$}{z,z1} \fmf{phantom_arrow,tension=1.5}{z,z1}
    \fmf{plain,tension=3,label.side=left,label=$q$}{x,x1} \fmf{phantom_arrow,tension=3,label.side=right,label=$e^0$}{x,x1}
    \fmf{plain,tension=3,label.side=left,label=$q+p$}{z1,p} \fmf{phantom_arrow,tension=3,label.side=right,label=$b^b$}{z1,p}
    \fmf{phantom,tension=3}{y,x1}
    \fmf{plain,label.side=right,label=$b^n$}{x1,xz} \fmf{plain,label.side=right,label=$e^m$}{xz,z1} \fmf{phantom_arrow,label.side=left,label=$k+q$}{x1,z1}
    \fmf{wiggly,right,label.side=right,label=$k$}{z1,x1} \fmf{phantom_arrow,right}{z1,x1}
    \fmfdot{x1,z1}
    \fmflabel{$\mu\, a$}{z}
    \end{fmfgraph*}
\end{fmffile}
\caption{Diagram for calculation of anticommuting fields
renormalization constant}\label{Fig_Z2Bar} \caption{Diagrams for
calculation of renormalization constant
$\bar{Z}_1$}\label{Fig_Z1Bar} }
\end{figure}

\section{Discussion.}
In the present paper it was shown that the theory with the gauge
invariant ghost Lagrangian agrees with the standard Yang-Mills theory in
one-loop approximation. The relationship $Z_1 Z_2^{-1} = \bar{Z_1}
\bar{Z_2}^{-1}$ between renormalization constants which is one of
the necessary conditions of gauge invariance of renormalized
Lagrangian holds. A complete renormalization of the theory
requires further investigation.

{\bf Acknowledgements} \\
I am grateful to A.A. Slavnov for useful discussion and remarks.

\newpage
\appendix
\section{Propagators and vertices.}
\begin{figure}[h]
\center{
\begin{fmffile}{graphs7}
    \begin{fmfgraph*}(50,40)
    \fmfleft{x} \fmfright{y}
    \fmf{wiggly,label.side=left,label=$A_\mu^a$}{x,xy} \fmf{wiggly,label.side=left,label=$A_\nu^b$}{xy,y} \fmf{phantom_arrow,label.side=right,label=$k$}{x,y}
    \fmfdot{x,y}
    \end{fmfgraph*}
    \begin{fmfgraph*}(90,40)
    \fmfleft{x} \fmfright{y}
    \fmfv{label.angle=0,label=$\dfrac{\delta^{ab}}{i}\dfrac{g_{\mu\nu}-k_\mu k_\nu/k^2}{k^2}$}{x}
    \end{fmfgraph*}
    \begin{fmfgraph}(55,40)
    \end{fmfgraph}
    \begin{fmfgraph*}(50,40)
    \fmfleft{x} \fmfright{y}
    \fmf{plain,label.side=left,label=$b^0$}{x,xy} \fmf{plain,label.side=left,label=$e^0$}{xy,y} \fmf{phantom_arrow,label.side=right,label=$k$}{x,y}
    \fmfdot{x,y}
    \end{fmfgraph*}
    \begin{fmfgraph*}(90,40)
    \fmfleft{x} \fmfright{y}
    \fmfv{label.angle=0,label=$\dfrac{1}{k^2}$}{x}
    \end{fmfgraph*}
    \begin{fmfgraph}(400,1)
    \end{fmfgraph}
    \begin{fmfgraph*}(50,40)
    \fmfleft{x} \fmfright{y}
    \fmf{wiggly,label.side=left,label=$A_\mu^a$}{x,xy} \fmf{plain,label.side=left,label=$B^{+b}$}{xy,y} \fmf{phantom_arrow,label.side=right,label=$k$}{x,y}
    \fmfdot{x,y}
    \end{fmfgraph*}
    \begin{fmfgraph*}(90,40)
    \fmfleft{x} \fmfright{y}
    \fmfv{label.angle=0,label=$\dfrac{\delta^{ab}}{i}\dfrac{\sqrt{2}}{a}\dfrac{(-ik_\mu)}{k^2}$}{x}
    \end{fmfgraph*}
    \begin{fmfgraph}(55,40)
    \end{fmfgraph}
    \begin{fmfgraph*}(50,40)
    \fmfleft{x} \fmfright{y}
    \fmf{plain,label.side=left,label=$b^a$}{x,xy} \fmf{plain,label.side=left,label=$e^b$}{xy,y} \fmf{phantom_arrow,label.side=right,label=$k$}{x,y}
    \fmfdot{x,y}
    \end{fmfgraph*}
    \begin{fmfgraph*}(90,40)
    \fmfleft{x} \fmfright{y}
    \fmfv{label.angle=0,label=$\dfrac{\delta^{ab}}{k^2}$}{x}
    \end{fmfgraph*}
    \begin{fmfgraph}(400,1)
    \end{fmfgraph}
    \begin{fmfgraph*}(50,40)
    \fmfleft{x} \fmfright{y}
    \fmf{dots,label.side=left,label=$\varphi^+$}{x,xy} \fmf{dots,label.side=left,label=$\varphi^-$}{xy,y} \fmf{phantom_arrow,label.side=right,label=$k$}{x,y}
    \fmfdot{x,y}
    \end{fmfgraph*}
    \begin{fmfgraph*}(90,40)
    \fmfleft{x} \fmfright{y}
    \fmfv{label.angle=0,label=$\dfrac{1}{i}\dfrac{(-2)}{k^2}$}{x}
    \end{fmfgraph*}
    \begin{fmfgraph}(55,40)
    \end{fmfgraph}
    \begin{fmfgraph*}(50,40)
    \end{fmfgraph*}
    \begin{fmfgraph*}(90,40)
    \end{fmfgraph*}
    \begin{fmfgraph}(400,8)
    \end{fmfgraph}
    \begin{fmfgraph*}(50,50)
    \fmftop{x} \fmfbottom{y,z}
    \fmf{wiggly}{c,x} \fmf{phantom_arrow,label.side=left,label=$k$}{c,x}
    \fmf{plain}{c,y} \fmf{phantom_arrow,label.side=left,label=$p$}{c,y}
    \fmf{dots}{c,z} \fmf{phantom_arrow,label.side=left,label=$q$}{c,z}
    \fmfdot{c}
    \fmflabel{$\mu\;a$}{x}
    \fmflabel{$B^{+b}$}{y}
    \fmflabel{$\varphi^-$}{z}
    \end{fmfgraph*}
    \begin{fmfgraph*}(90,50)
    \fmfleft{x}
    \fmfv{label.angle=0,label=$\dfrac{ig}{4}\delta^{ab}i(p-q)_\mu$}{x}
    \end{fmfgraph*}
    \begin{fmfgraph}(55,50)
    \end{fmfgraph}
    \begin{fmfgraph*}(50,50)
    \fmftop{x,y} \fmfbottom{z}
    \fmf{wiggly,tension=1.5}{x,c}
    \fmf{wiggly,tension=1.5}{y,c}
    \fmf{dots,label.side=left,label=$\varphi^+$}{c,z} \fmf{phantom_arrow}{c,z}
    \fmfdot{c}
    \fmflabel{$\mu\;a$}{x}
    \fmflabel{$\nu\;b$}{y}
    \end{fmfgraph*}
    \begin{fmfgraph*}(90,50)
    \fmfleft{x}
    \fmfv{label.angle=0,label=$\dfrac{iag\sqrt{2}}{2} g_{\mu\nu}\delta^{ab}$}{x}
    \end{fmfgraph*}
    \begin{fmfgraph}(400,8)
    \end{fmfgraph}
    \begin{fmfgraph*}(50,50)
    \fmftop{x,y} \fmfbottom{z,m}
    \fmf{wiggly}{c,x}
    \fmf{wiggly}{c,y}
    \fmf{dots_arrow}{c,z}
    \fmf{dots_arrow}{c,m}
    \fmfdot{c}
    \fmflabel{$\mu\;a$}{x}
    \fmflabel{$\nu\;b$}{y}
    \fmflabel{$\varphi^+$}{z}
    \fmflabel{$\varphi^-$}{m}
    \end{fmfgraph*}
    \begin{fmfgraph*}(90,50)
    \fmfleft{x}
    \fmfv{label.angle=0,label=$\dfrac{ig^2}{4}g_{\mu\nu}\delta^{ab}$}{x}
    \end{fmfgraph*}
    \begin{fmfgraph}(55,50)
    \end{fmfgraph}
    \begin{fmfgraph*}(50,50)
    \fmftop{x} \fmfbottom{y,z}
    \fmf{plain_arrow,label.side=left,label=$p$}{c,x}
    \fmf{wiggly,label.side=left,label=$k$}{y,c}
    \fmf{plain_arrow,label.side=left,label=$q$}{z,c}
    \fmfdot{c}
    \fmflabel{$b^b$}{x}
    \fmflabel{$\mu\;a$}{y}
    \fmflabel{$e^c$}{z}
    \end{fmfgraph*}
    \begin{fmfgraph*}(90,50)
    \fmfleft{x}
    \fmfv{label.angle=0,label=$\dfrac{-ig}{2}\varepsilon^{abc}(p+q)_\mu$}{x}
    \end{fmfgraph*}
    \begin{fmfgraph}(400,8)
    \end{fmfgraph}
    \begin{fmfgraph*}(50,50)
    \fmftop{x} \fmfbottom{y,z}
    \fmf{plain_arrow,label.side=left,label=$p$}{c,x}
    \fmf{wiggly,label.side=left,label=$k$}{y,c}
    \fmf{plain_arrow,label.side=left,label=$q$}{z,c}
    \fmfdot{c}
    \fmflabel{$b^b$}{x}
    \fmflabel{$\mu\;a$}{y}
    \fmflabel{$e^0$}{z}
    \end{fmfgraph*}
    \begin{fmfgraph*}(90,50)
    \fmfleft{x}
    \fmfv{label.angle=0,label=$\dfrac{-ig}{2}\delta^{ab}(p+q)_\mu$}{x}
    \end{fmfgraph*}
    \begin{fmfgraph}(55,50)
    \end{fmfgraph}
    \begin{fmfgraph*}(50,50)
    \fmftop{x} \fmfbottom{y,z}
    \fmf{plain_arrow,label.side=left,label=$p$}{c,x}
    \fmf{wiggly,label.side=left,label=$k$}{y,c}
    \fmf{plain_arrow,label.side=left,label=$q$}{z,c}
    \fmfdot{c}
    \fmflabel{$b^0$}{x}
    \fmflabel{$\mu\;a$}{y}
    \fmflabel{$e^c$}{z}
    \end{fmfgraph*}
    \begin{fmfgraph*}(90,50)
    \fmfleft{x}
    \fmfv{label.angle=0,label=$\dfrac{ig}{2}\delta^{ab}(p+q)_\mu$}{x}
    \end{fmfgraph*}
    \begin{fmfgraph}(400,8)
    \end{fmfgraph}
    \begin{fmfgraph*}(50,50)
    \fmftop{x,y} \fmfbottom{z,m}
    \fmf{wiggly}{c,x}
    \fmf{wiggly}{c,y}
    \fmf{plain_arrow}{z,c}
    \fmf{plain_arrow}{c,m}
    \fmfdot{c}
    \fmflabel{$\mu\;a$}{x}
    \fmflabel{$\nu\;b$}{y}
    \fmflabel{$e^c$}{z}
    \fmflabel{$b^d$}{m}
    \end{fmfgraph*}
    \begin{fmfgraph*}(90,50)
    \fmfleft{x}
    \fmfv{label.angle=0,label=$-\dfrac{g^2}{2}g_{\mu\nu}\delta^{ab}\delta^{cd}$}{x}
    \end{fmfgraph*}
    \begin{fmfgraph}(55,50)
    \end{fmfgraph}
    \begin{fmfgraph*}(50,50)
    \fmftop{x,y} \fmfbottom{z,m}
    \fmf{wiggly}{c,x}
    \fmf{wiggly}{c,y}
    \fmf{plain_arrow}{z,c}
    \fmf{plain_arrow}{c,m}
    \fmfdot{c}
    \fmflabel{$\mu\;a$}{x}
    \fmflabel{$\nu\;b$}{y}
    \fmflabel{$e^0$}{z}
    \fmflabel{$b^0$}{m}
    \end{fmfgraph*}
    \begin{fmfgraph*}(90,50)
    \fmfleft{x}
    \fmfv{label.angle=0,label=$-\dfrac{g^2}{2}g_{\mu\nu}\delta^{ab}$}{x}
    \end{fmfgraph*}
    \begin{fmfgraph}(400,8)
    \end{fmfgraph}
\end{fmffile}
}
\end{figure}

\section{Calculation of $Z_2$.}
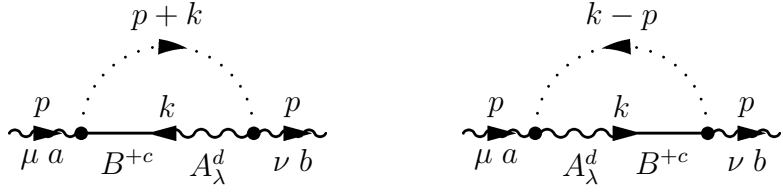
\begin{figure}
\center{
\begin{fmffile}{graphs1}
    \begin{fmfgraph*}(120,100)
    \fmfleft{x} \fmfright{y}
    \fmf{wiggly,tension=5,label.side=left,label=$p$}{x,x1} \fmf{phantom_arrow,label.side=right,label=$\mu\; a$}{x,x1}
    \fmf{wiggly,tension=5,label.side=left,label=$p$}{y1,y} \fmf{phantom_arrow,label.side=right,label=$\nu\; b$}{y1,y}
    \fmf{dots_arrow,left,label.side=left,label=$p+k$}{x1,y1}
    \fmf{wiggly,label.side=left,label=$A_\lambda^d$}{y1,z}
    \fmf{plain,label.side=left,label=$B^{+c}$}{z,x1}
    \fmf{phantom_arrow,label.side=right,label=$k$}{y1,x1}
    \fmfdot{x1,y1}
    \end{fmfgraph*}
    \begin{fmfgraph}(40,100)
    \end{fmfgraph}
    \begin{fmfgraph*}(120,100)
    \fmfleft{x} \fmfright{y}
    \fmf{wiggly,tension=5,label.side=left,label=$p$}{x,x1} \fmf{phantom_arrow,label.side=right,label=$\mu\; a$}{x,x1}
    \fmf{wiggly,tension=5,label.side=left,label=$p$}{y1,y} \fmf{phantom_arrow,label.side=right,label=$\nu\; b$}{y1,y}
    \fmf{dots_arrow,right,label.side=right,label=$k-p$}{y1,x1}
    \fmf{wiggly,label.side=right,label=$A_\lambda^d$}{x1,z}
    \fmf{plain,label.side=right,label=$B^{+c}$}{z,y1}
    \fmf{phantom_arrow,label.side=left,label=$k$}{x1,y1}
    \fmfdot{x1,y1}
    \end{fmfgraph*}
\end{fmffile}
\caption{1PI diagrams with loops of commuting fields $B^+$,
$\varphi^\pm$}\label{Fig_Z2Commute} }
\end{figure}
Here we present the calculations of the commuting scalar fields~$B^{+,a}$,
$\varphi$ and anticommuting fields~$e$, $b$ loops contributions
to the divergent part of the gauge field two-point Green function.
\par For calculation of commuting scalar fields contribution it is
necessary to find divergent part of two diagrams which are
represented at the fig.~\ref{Fig_Z2Commute}. Left diagram~(fig.
\ref{Fig_Z2Commute}) gives the following integral.
\begin{multline*}
    \int \frac{d^4k}{(2\pi)^4} \left( -\frac{g}{4} \delta^{ac}
    (-2k-p)_\mu \right) \frac{1}{i} \frac{(-2)}{(p+k)^2} \left(
    \frac{iag\sqrt{2}}{2} g_{\lambda\nu} \delta^{db} \right)
    \delta^{cd} \frac{\sqrt{2}}{a} \frac{(-k_\lambda)}{k^2} =\\=
    \frac{g^2}{2} \delta^{ab} \int \frac{d^4k}{(2\pi)^4} \frac{(2k+p)_\mu k_\nu}{k^2
    (p+k)^2} \stackrel{div}{=} \frac{g^2}{2} \delta^{ab}
    \frac{(-i)}{48\pi^2\varepsilon} (g_{\mu\nu} p^2 - p_\mu p_\nu)
\end{multline*}
It is easy to see that the right diagram at fig.~\ref{Fig_Z2Commute}
coincides with the left diagram after the change
$\mu\leftrightarrow\nu$, $a\leftrightarrow b$, $p\rightarrow -p$
and consequently it gives the same result. Therefore the total
commuting scalar fields contribution is
\begin{equation}\label{App_Z2_commute}
    -\frac{ig^2}{48\pi^2\varepsilon} \delta^{ab} (g_{\mu\nu} p^2 -
    p_\mu p_\nu)
\end{equation}
\begin{figure}[b]
\center{
\begin{fmffile}{graphs2}
    \begin{fmfgraph*}(120,100)
    \fmfleft{x} \fmfright{y}
    \fmf{wiggly,tension=3,label.side=left,label=$p$}{x,x1} \fmf{phantom_arrow,tension=3,label.side=right,label=$\mu\; a$}{x,x1}
    \fmf{wiggly,tension=3,label.side=left,label=$p$}{y1,y} \fmf{phantom_arrow,tension=3,label.side=right,label=$\nu\; b$}{y1,y}
    \fmf{plain_arrow,left,label.side=left,label=$b^m\qquad e^n$}{x1,y1} \fmf{plain,left,label.side=right,label=$k$}{x1,y1}
    \fmf{plain_arrow,left,label.side=left,label=$e^c\qquad b^l$}{y1,x1} \fmf{plain,left,label.side=right,label=$k-p$}{y1,x1}
    \fmfdot{x1,y1}
    \end{fmfgraph*}
    \begin{fmfgraph}(10,100)
    \end{fmfgraph}
    \begin{fmfgraph*}(120,100)
    \fmfleft{x} \fmfright{y}
    \fmf{wiggly,tension=3}{x,x1} \fmf{phantom_arrow,tension=3}{x,x1}
    \fmf{wiggly,tension=3}{y1,y} \fmf{phantom_arrow,tension=3}{y1,y}
    \fmf{plain_arrow,left,label.side=left,label=$b^0\qquad e^0$}{x1,y1} \fmf{plain,left}{x1,y1}
    \fmf{plain_arrow,left,label.side=left,label=$e^c\qquad b^l$}{y1,x1} \fmf{plain,left}{y1,x1}
    \fmfdot{x1,y1}
    \end{fmfgraph*}
    \begin{fmfgraph}(10,100)
    \end{fmfgraph}
    \begin{fmfgraph*}(120,100)
    \fmfleft{x} \fmfright{y}
    \fmf{wiggly,tension=3}{x,x1} \fmf{phantom_arrow,tension=3}{x,x1}
    \fmf{wiggly,tension=3}{y1,y} \fmf{phantom_arrow,tension=3}{y1,y}
    \fmf{plain_arrow,left,label.side=left,label=$b^m\qquad e^n$}{x1,y1} \fmf{plain,left}{x1,y1}
    \fmf{plain_arrow,left,label.side=left,label=$e^0\qquad b^0$}{y1,x1} \fmf{plain,left}{y1,x1}
    \fmfdot{x1,y1}
    \end{fmfgraph*}
\end{fmffile}
\caption{1PI diagrams with loops of anticommuting
fields}\label{Fig_Z2Anticommute} }
\end{figure}
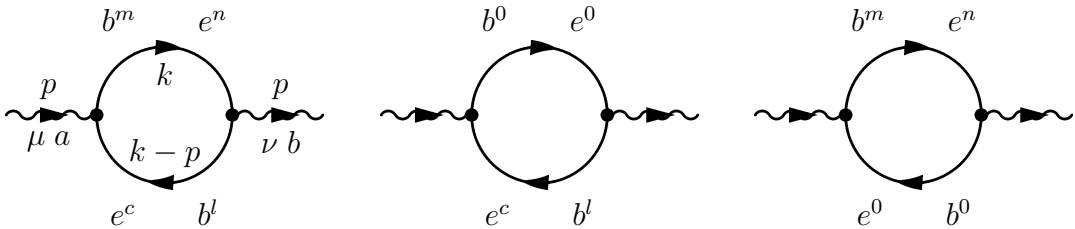
\par Now let us find the anticommuting fields contribution. We calculate
divergent part of diagrams shown at fig.~\ref{Fig_Z2Anticommute}.
For the first diagram (fig.~\ref{Fig_Z2Anticommute}) we have
\begin{multline*}
    (-1) \int \frac{d^4k}{(2\pi)^4} \left( -\frac{ig}{2}
    \varepsilon^{amc} (2k-p)_\mu \right) \frac{\delta^{mn}}{k^2}
    \left( -\frac{ig}{2} \varepsilon^{bln} (2k-p)_\nu \right)
    \frac{\delta^{lc}}{(k-p)^2} =\\= \frac{g^2}{4} \varepsilon^{anl}
    \varepsilon^{bln} \int \frac{d^4k}{(2\pi)^4}
    \frac{(2k-p)_\mu(2k-p)_\nu}{k^2(k-p)^2} \stackrel{div}{=}
    \frac{ig^2}{48\pi^2\varepsilon} \delta^{ab} (g_{\mu\nu}p^2 - p_\mu p_\nu)
\end{multline*}
One can see that the remaining diagrams have identical divergent parts.
Furthermore sum of these divergences is equal to the divergence of the
first diagram. Therefore the total anticommuting fields contribution
is
\begin{equation}\label{App_Z2_anticommute}
    \frac{ig^2}{24\pi^2\varepsilon} \delta^{ab} (g_{\mu\nu}p^2 - p_\mu p_\nu)
\end{equation}
Summarizing~(\ref{App_Z2_commute}) and~(\ref{App_Z2_anticommute})
we find
\begin{equation}\label{App_Z2_sumCommAnticomm}
    \frac{ig^2}{48\pi^2\varepsilon} \delta^{ab} (g_{\mu\nu}p^2 - p_\mu p_\nu)
\end{equation}

\section{Calculation of $Z_1$.}
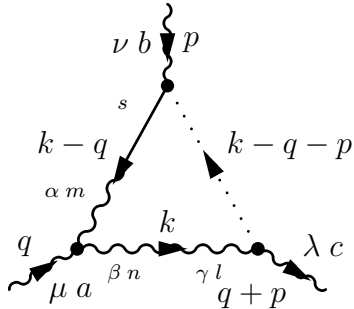
\begin{figure}[b]
\center{
\begin{fmffile}{graphs3}
    \begin{fmfgraph*}(120,120)
    \fmftop{x} \fmfbottom{y,z}
    \fmf{wiggly,tension=3,label.side=left,label=$p$}{x,x1} \fmf{phantom_arrow,tension=3,label.side=right,label=$\nu\; b$}{x,x1}
    \fmf{wiggly,tension=3,label.side=left,label=$q$}{y,y1} \fmf{phantom_arrow,tension=3,label.side=right,label=$\mu\; a$}{y,y1}
    \fmf{wiggly,tension=3,label.side=right,label=$q+p$}{z1,z} \fmf{phantom_arrow,tension=3,label.side=left,label=$\lambda\; c$}{z1,z}
    \fmf{plain,label.side=right,label=${}^s$}{x1,xy} \fmf{wiggly,label.side=right,label=${}_{\alpha\; m}$}{xy,y1} \fmf{phantom_arrow,label.side=right,label=$k-q$}{x1,y1}
    \fmf{wiggly,label.side=right,label=${}^{\beta\; n}$}{y1,yz} \fmf{wiggly,label.side=right,label=${}^{\gamma\; l}$}{yz,z1} \fmf{phantom_arrow,label.side=left,label=$k$}{y1,z1}
    \fmf{dots}{z1,xz} \fmf{dots}{xz,x1} \fmf{phantom_arrow,label.side=right,label=$k-q-p$}{z1,x1}
    \fmfdot{x1,y1,z1}
    \end{fmfgraph*}
\end{fmffile}
\caption{Diagram with commuting fields loop}\label{Fig_Z1Commute}
}
\end{figure}
Now we calculate the contribution of the commuting scalar
fields~$B^{+,a}$, $\varphi$ and the anticommuting fields~$e$, $b$
into divergent part of the  gauge field three-point Green
function.
\par The contribution of commuting fields~$B^{+,a}$ and
$\varphi^\pm$ loops is given by the sum of six diagrams one of which is
represented at fig.~\ref{Fig_Z1Commute} and other five diagrams
differ from this one by permutation of vertices. The diagram
at fig.~\ref{Fig_Z1Commute} gives
\begin{multline}\label{Z1_commute1of6}
    \int \frac{d^4k}{(2\pi)^4} \left( -g \varepsilon^{amn} \{
    (k-2q)_\beta g_{\mu\alpha} + (q-2k)_\mu g_{\alpha\beta} +
    (k+q)_\alpha g_{\beta\mu} \} \right) \frac{\delta^{nl}}{i}
    \frac{g_{\beta\gamma}-k_\beta k_\gamma /k^2}{k^2} \\ \left(
    \frac{iag\sqrt{2}}{2} g_{\gamma\lambda} \delta^{lc} \right)
    \frac{1}{i} \frac{(-2)}{(k-q-p)^2} \left( -\frac{g}{4}
    \delta^{ls} (2k-2q-p)_\nu \right) \delta^{ms}
    \frac{\sqrt{2}}{a} \frac{(k-q)_\alpha}{(k-q)^2} \stackrel{div}{=}\\\stackrel{div}{=}
    -\frac{g^3}{16\pi^2\varepsilon} \varepsilon^{abc} \left(
    -\frac{11}{96} g_{\mu\nu} (p+2q)_\lambda - \frac{11}{96} g_{\nu\lambda}
    (p+2q)_\mu + \frac{13}{96} g_{\mu\lambda} (p+2q)_\nu \right)
\end{multline}
Contributions of five remaining diagrams can be obtained from
formula~(\ref{Z1_commute1of6}). For example if we replace $q
\rightarrow (-q-p)$, $p \rightarrow p$, $a\leftrightarrow c$, $\mu
\leftrightarrow \lambda$ in this formula, we obtain the divergent part
of the diagram which differs by permutation of two bottom vertices.
Summarizing all six we obtain
\begin{equation}\label{Z1_commute}
    \left. \left. -\frac{g^3}{32\pi^2\varepsilon} \varepsilon^{abc}
    \right( g_{\mu\nu} (p-q)_\lambda - g_{\nu\lambda} (2p+q)_\mu +
    g_{\mu\lambda} (p+2q)_\nu \right)
\end{equation}
\begin{figure}
\center{
\begin{fmffile}{graphs4}
    \begin{fmfgraph*}(120,120)
    \fmftop{x} \fmfbottom{y,z}
    \fmf{wiggly,tension=3,label.side=left,label=$p$}{x,x1} \fmf{phantom_arrow,tension=3,label.side=right,label=$\nu\; b$}{x,x1}
    \fmf{wiggly,tension=3,label.side=left,label=$q$}{y,y1} \fmf{phantom_arrow,tension=3,label.side=right,label=$\mu\; a$}{y,y1}
    \fmf{wiggly,tension=3,label.side=right,label=$q+p$}{z1,z} \fmf{phantom_arrow,tension=3,label.side=left,label=$\lambda\; c$}{z1,z}
    \fmf{plain,label.side=right,label=$e^f$}{x1,xy} \fmf{plain,label.side=right,label=$b^r$}{xy,y1} \fmf{phantom_arrow,label.side=left,label=$k+q$}{y1,x1}
    \fmf{plain,label.side=right,label=$e^0$}{y1,yz} \fmf{plain,label.side=right,label=$b^0$}{yz,z1} \fmf{phantom_arrow,label.side=right,label=$k$}{z1,y1}
    \fmf{plain,label.side=right,label=$e^n$}{z1,xz} \fmf{plain,label.side=right,label=$b^m$}{xz,x1} \fmf{phantom_arrow,label.side=left,label=$k+p+q$}{x1,z1}
    \fmfdot{x1,y1,z1}
    \end{fmfgraph*}
    \begin{fmfgraph}(30,120)
    \end{fmfgraph}
    \begin{fmfgraph*}(200,120)
    \fmfleft{x} \fmfright{y}
    \fmf{phantom,label.side=left,label={$+$ five diagrams differing by}}{x,y}
    \fmf{phantom,label.side=right,label={permutation of vertices}}{x,y}
    \end{fmfgraph*}
    \begin{fmfgraph*}(400,10)
    \end{fmfgraph*}
    \begin{fmfgraph*}(120,120)
    \fmftop{x} \fmfbottom{y,z}
    \fmf{wiggly,tension=3}{x,x1} \fmf{phantom_arrow,tension=3}{x,x1}
    \fmf{wiggly,tension=3}{y,y1} \fmf{phantom_arrow,tension=3}{y,y1}
    \fmf{wiggly,tension=3}{z1,z} \fmf{phantom_arrow,tension=3}{z1,z}
    \fmf{plain,label.side=right,label=$e^f$}{x1,xy} \fmf{plain,label.side=right,label=$b^r$}{xy,y1} \fmf{phantom_arrow}{y1,x1}
    \fmf{plain,label.side=right,label=$e^s$}{y1,yz} \fmf{plain,label.side=right,label=$b^l$}{yz,z1} \fmf{phantom_arrow}{z1,y1}
    \fmf{plain,label.side=right,label=$e^n$}{z1,xz} \fmf{plain,label.side=right,label=$b^m$}{xz,x1} \fmf{phantom_arrow}{x1,z1}
    \fmfdot{x1,y1,z1}
    \end{fmfgraph*}
    \begin{fmfgraph}(30,120)
    \end{fmfgraph}
    \begin{fmfgraph*}(120,120)
    \fmftop{x} \fmfbottom{y,z}
    \fmf{wiggly,tension=3}{x,x1} \fmf{phantom_arrow,tension=3}{x,x1}
    \fmf{wiggly,tension=3}{y,y1} \fmf{phantom_arrow,tension=3}{y,y1}
    \fmf{wiggly,tension=3}{z1,z} \fmf{phantom_arrow,tension=3}{z1,z}
    \fmf{plain}{x1,xy} \fmf{plain}{xy,y1} \fmf{phantom_arrow}{x1,y1}
    \fmf{plain}{y1,yz} \fmf{plain}{yz,z1} \fmf{phantom_arrow}{y1,z1}
    \fmf{plain}{z1,xz} \fmf{plain}{xz,x1} \fmf{phantom_arrow}{z1,x1}
    \fmfdot{x1,y1,z1}
    \end{fmfgraph*}
\end{fmffile}
\caption{Diagrams with anticommuting fields
loops}\label{Fig_Z1Anticommute} }
\end{figure}
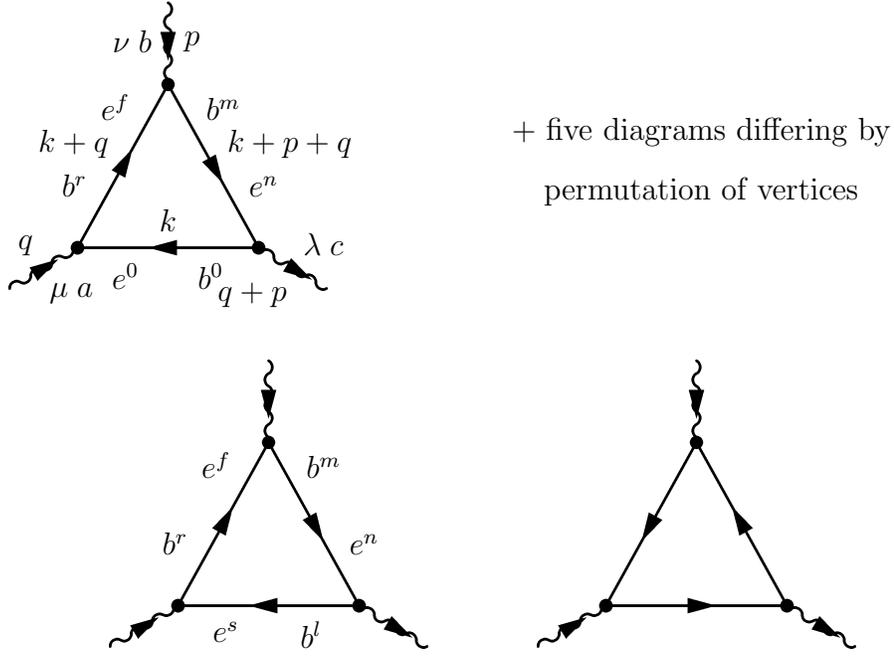
\par Now let us consider the loops of the anticommuting fields~$e$ and $b$.
Corresponding diagrams are presented at fig.~\ref{Fig_Z1Anticommute}. Vertices $A_\mu^a e^m b^n$, $A_\mu^a
e^0 b^n$, $A_\mu^a e^m b^0$ differ from each other only by tensors
$\varepsilon^{amn}$, $\delta^{an}$, $-\delta^{am}$
correspondingly.  For the top and bottom diagrams we obtain correspondingly
\begin{gather*}
    \delta^{ar} \delta^{rf} \varepsilon^{bmf} \delta^{mn}
    (-\delta^{nc}) = -\varepsilon^{bca} = -\varepsilon^{abc}\\
    \varepsilon^{ars} \delta^{rf} \varepsilon^{bmf} \delta^{mn}
    \varepsilon^{cln} \delta^{ls} = \varepsilon^{afs}
    \varepsilon^{bnf} \varepsilon^{csn} = \varepsilon^{afs} (-1)
    (\delta^{bc} \delta^{fs} - \delta^{bs} \delta^{fc}) = -\varepsilon^{abc}
\end{gather*}
So these diagrams give the same results. In addition one can see
that each remaining diagrams give the same result too. So we have
to calculate divergent part of the first diagram at
fig.~\ref{Fig_Z1Anticommute} and multiply it by eight.
\begin{multline*}
    -\varepsilon^{abc} \int \frac{d^4k}{(2\pi)^4} \left(
    -\frac{ig}{2} (2k+q)_\mu \right) \frac{1}{k^2} \left(
    -\frac{ig}{2} (2k+p+q)_\lambda \right) \frac{1}{(k+p+q)^2}\\ \left(
    -\frac{ig}{2} (2k+2q+p)_\nu \right) \frac{1}{(k+q)^2} \stackrel{div}{=}\\\stackrel{div}{=}
    \left. \left. -\frac{ig^3}{8} \varepsilon^{abc} \frac{i}{24\pi^2\varepsilon}
    \right( g_{\mu\nu} (p-q)_\lambda - g_{\nu\lambda} (2p+q)_\mu +
    g_{\mu\lambda} (p+2q)_\nu \right)
\end{multline*}
 The total contribution of the anticommuting fields loops and
loops of the fields $B^{+,a}$, $\varphi^\pm$ is
\begin{equation}\label{Z1_sumCommAnticomm}
    \left.\left. \frac{g^3}{96\pi^2\varepsilon} \varepsilon^{abc}
    \right( g_{\mu\nu} (p-q)_\lambda - g_{\nu\lambda} (2p+q)_\mu +
    g_{\mu\lambda} (p+2q)_\nu \right).
\end{equation}

\section{Anticommuting fields $e$ and $b$.}
Now we calculate the constants $\bar{Z_2}$, $\bar{Z_1}$. Let us
consider the one-loop diagram at fig.~\ref{Fig_Z2Bar}. The corresponding
integral is equal to
\begin{multline*}
    \int \frac{d^4k}{(2\pi)^4} \left( -\frac{ig}{2} \delta^{am}
    (2p+k)_\mu \right) \frac{\delta^{mn}}{(p+k)^2} \left( \frac{ig}{2} \delta^{bn}
    (2p+k)_\nu \right) \frac{\delta^{ab}}{i} \frac{g_{\mu\nu}-k_\mu
    k_\nu/k^2}{k^2} = \\ = -3ig^2 p_\mu p_\nu \int
    \frac{d^4k}{(2\pi)^4} \frac{g_{\mu\nu}k^2 -k_\mu k_\nu}{k^4(p+k)^2}
    \stackrel{div}{=}
    \frac{9g^2}{32\pi^2\varepsilon} p^2
\end{multline*}
This divergence is canceled by the following counterterm:
\begin{gather}
    i \int dx (\bar{Z_2}-1) (i \partial_\mu e^0 \partial_\mu b^0)
    = \int dp (\bar{Z_2}-1) (-p^2) e^0(p) b^0(-p) \notag\\
    \mbox{thus\qquad}-(\bar{Z_2}-1)p^2 = -\frac{9g^2}{32\pi^2\varepsilon} p^2 \notag\\
    \mbox{and therefore\qquad}\bar{Z_2} = 1 +
    \frac{9g^2}{32\pi^2\varepsilon} \label{Z2bar}
\end{gather}
To find the constant $\bar{Z}_1$ one has to calculate the sum of
divergent parts of diagrams which are presented at
fig.~\ref{Fig_Z1Bar}.
 Besides the presented diagrams
there are the diagrams  with three vertices, but their divergent parts are
equal to zero.
\par For the left diagram at fig.~\ref{Fig_Z1Bar} we obtain the
following integral.
\begin{multline}\label{Z1bar_left}
    \int \frac{d^4k}{(2\pi)^4} \left( -\frac{g^2}{2} g_{\mu\nu}
    \delta^{ac} \right) \frac{1}{(k+p+q)^2} \left( -\frac{ig}{2}
    \delta^{bd} (k+2p+2q)_\lambda \right) \frac{\delta^{cd}}{i}
    \frac{g_{\nu\lambda} - k_\nu k_\lambda /k^2}{k^2} = \\ =
    \frac{g^3}{2} \delta^{ab} (p+q)_\lambda \int
    \frac{d^4k}{(2\pi)^4} \frac{g_{\mu\nu}k^2 -k_\mu k_\nu}{k^4(k+p+q)^2}
    \stackrel{div}{=} \frac{g^3}{2} \delta^{ab} (p+q)_\mu
    \frac{3i}{32\pi^2\varepsilon}
\end{multline}
For the right diagram on fig.~\ref{Fig_Z1Bar} we obtain
\begin{multline}\label{Z1bar_right}
    \int \frac{d^4k}{(2\pi)^4} \left( -\frac{ig}{2}
    \delta^{cn} (k+2q)_\nu \right) \frac{\delta^{nm}}{(k+q)^2} \left( -\frac{g^2}{2}
    g_{\mu\lambda} \delta^{ad} \delta^{mb} \right) \frac{\delta^{cd}}{i}
    \frac{g_{\nu\lambda} - k_\nu k_\lambda /k^2}{k^2} = \\ =
    \frac{g^3}{2} \delta^{ab} q_\nu \int
    \frac{d^4k}{(2\pi)^4} \frac{g_{\mu\nu}k^2 -k_\mu k_\nu}{k^4(k+q)^2}
    \stackrel{div}{=} \frac{g^3}{2} \delta^{ab} q_\mu
    \frac{3i}{32\pi^2\varepsilon}
\end{multline}
Summarizing the results~(\ref{Z1bar_left}) and~(\ref{Z1bar_right})
one gets
\begin{equation}
    -\frac{ig^3}{2} \delta^{ab} (p+2q)_\mu \frac{(-3)}{32\pi^2\varepsilon}
\end{equation}
This divergencies are canceled by counterterm $(\bar{Z_1}-1)
\frac{(-ig)}{2} \delta^{ab} (p+2q)_\mu$. So
\begin{equation}\label{Z1bar}
    \bar{Z_1} = 1 + \frac{3g^2}{32\pi^2\varepsilon}
\end{equation}

\end{document}